
\input vanilla.sty
\input harvmac

\newif\iffigs\figstrue

  \iffigs
   \input epsf
\else
   \message{No figures will be included. See TeX file for more
information}
\fi

\def\inbar{\,\vrule height1.5ex width.4pt depth0pt}
\font\cmss=cmss10 \font\cmsss=cmss8 at 8pt
\def\BZ{\relax\ifmmode\mathchoice
{\hbox{\cmss Z\kern-.4em Z}}{\hbox{\cmss Z\kern-.4em Z}}
{\lower.9pt\hbox{\cmsss Z\kern-.36em Z}}
{\lower1.2pt\hbox{\cmsss Z\kern-.36em Z}}\else{\cmss Z\kern-.4em Z}\fi}
\def\IC{\relax\hbox{$\inbar\kern-.3em\hbox{\rm C}$}}
\def\IP{\relax\hbox{\rm I\kern-.18em P}}
\def\IQ{\relax\hbox{$\inbar\kern-.3em\hbox{\rm Q}$}}
\def\IR{\relax\hbox{\rm I\kern-.18em R}}
\let\IZ\BZ
\let\ICP\IP
\def\F{\hbox{\it F}}
\def\wgt4 #1 #2 #3 #4 #5{\IP^4_{#1,#2,#3,#4,#5}}
\font\sevenrm=cmr7


 1
 1
 1

\font\tenbf=cmbx10
\font\tenrm=cmr10
\font\tenit=cmti10

\font\eightrm=cmr8
\font\eightit=cmti8

\parindent=1.2pc
\magnification=\magstep1
\hsize=6.0truein
\vsize=8.6truein

\rightline{hep-th/9509049, CLNS-95/1361}
\vglue0.3cm

\centerline{\tenbf PHASES OF MIRROR SYMMETRY}
\baselineskip=18pt
\centerline{\eightrm TI-MING CHIANG and BRIAN R. GREENE\raise 1ex \hbox{\dag}}
\baselineskip=18pt
\vfootnote \dag{\eightrm  Talk delivered by B.R.G at the {\it Strings '95}
conference, March 13-18, 1995, USC.}
\baselineskip=12pt
\centerline{\eightit F.R. Newman Lab. of Nuclear Studies, Cornell University}
\baselineskip=10pt
\centerline{\eightit Ithaca, NY 14853, USA}

\vglue0.6cm
\centerline{\bf Abstract}
\vglue0.2cm

{\rightskip=3pc
\leftskip=3pc
\eightrm\baselineskip=10pt\noindent
We review the geometrical framework required for understanding the moduli
space of $(2,2)$ superconformal-field theories, highlighting various
aspects of its phase structure. In particular, we indicate the types of phase
diagrams that emerge for ``generic'' Calabi-Yau theories and review an
efficient method for their determination. We then focus on some special
types of phase diagrams that have bearing on the issues of rigid
manifolds, mirror symmetry and geometrical duality.
\vglue0.6cm}

\tenrm\baselineskip=13pt
\leftline{\tenbf 1. Introduction}
\vglue0.4cm

In the last few years a tremendous amount of progress has been made
in understanding the structure of $N = 2$ conformal field theory
moduli space. Most impressive has been the emergence of a rich phase structure
\ref\rWittenphases{E. Witten, {\it Phases of N=2 Theories in Two
Dimensions}, Nucl. Phys. {\bf B403} (1993) 159.}\
\ref\rAGM{P. Aspinwall, B. Greene, and D. Morrison, {\it Calabi-Yau Moduli
Space, Mirror Manifolds, and Spacetime Topology Change in String
Theory}, Nucl. Phys. {\bf B416} (1994) 414.}\ for those theories that
can be realized as the infrared limit of $N = 2$ two dimensional gauged
linear sigma models \rWittenphases. Amongst such theories, for the case of
abelian gauge groups, are nonlinear sigma models on Calabi-Yau target
spaces and Landau-Ginzburg theories. The phase structure describing these
theories has led both to interesting new physics, such as spacetime
topology change \rWittenphases\ \rAGM, as well as clarified some
previously puzzling features of mirror symmetry. Examples of the latter
include the precise statement of the relationship between the K\"ahler
moduli space of a Calabi-Yau space and the complex structure moduli space
of its mirror (which is closely related to the results on spacetime
topology change) as well as the resolution of how mirror symmetry applies
to rigid Calabi-Yau manifolds (whose putative mirror would appear to lack
a K\"ahler form).

In the present talk we will briefly review the phase picture of $N = 2$
moduli space, discuss its application to ``generic'' examples, and then
illustrate its utility in discussing ``non-generic'' examples. Among the
generic examples we illustrate the tremendous growth rate of the number
of phases with increasing $h^{1,1}$. For instance, we note an example
with more than 700,000 phases. Among the nongeneric examples are cases in
which there's no Calabi-Yau sigma model phase, no Landau-Ginzburg phase,
or {\it all} Calabi-Yau sigma model phases. We discuss these examples
with particular reference to the issues of mirrors of rigid manifolds and
geometrical $R \leftrightarrow 1/R$ type duality.

\vglue0.6cm
\leftline{\tenbf 2. The Phase Picture}
\vglue0.4cm

The phase picture of $N = 2$ conformal field theory moduli space was
arrived at by two independent lines of investigation \rWittenphases\
\rAGM. The former sought to provide a more robust argument
relating Calabi-Yau sigma models to Landau-Ginzburg theories than
provided in \ref\rGVW{ B.Greene, C. Vafa and N. Warner, {\it Calabi-Yau
Manifolds and Renormalization Group Flows}, Nucl. Phys. {\bf B324}
(1989) 371.} while the latter sought to resolve a troubling issue in
mirror symmetry. When the dust settles, the two analyses yield the same
result: the classical K\"ahler moduli space of a Calabi-Yau sigma model
is but a subset of the corresponding quantum conformal field theory
moduli space. The latter ``enlarged K\"ahler moduli space'' consists of
numerous regions glued along common boundaries with the natural identity
of the corresponding physical model changing from region to region. By
natural identity we mean the formulation of the model for which
perturbation theory is reliable. In the simplest examples of one
dimensional moduli spaces there are two regions -- typically consisting
of a Calabi-Yau sigma model region and a Landau-Ginzburg region. Examples
with higher dimensional moduli spaces can have a far richer phase
structure.

We first review the phase structure of these moduli spaces from the viewpoint
of \rAGM\ and subsequently return to the approach of \rWittenphases\ by
establishing a dictionary between these two approaches, as discussed in
\ref\rAG{P. Aspinwall and B. Greene, {\it On the Geometric
Interpretation of N = 2 Superconformal Theories}, hep-th/9409110}.

The problem which motivated the work in \rAGM\ is simple to state: according to
mirror symmetry if $M$ and $W$ are a mirror pair, then the complex structure
moduli space of $M$ isomorphic to the (complexified) K\"ahler moduli space of
$W$ and vice versa; mathematically, though, they appear to be anything but
isomorphic. This problem is most easily understood by directly comparing
the typical form of a complexified K\"ahler moduli space with that of a complex
structure moduli space.

\vglue0.6cm
\leftline{\tenit 2.1 K\"ahler Moduli Space}
\vglue0.4cm

Classically, the K\"ahler form on a Calabi-Yau space is a closed two form $J$
related to the metric $g$ via
\eqn\eKahler{J = i g_{i \overline j}dX^i \wedge dX^{\overline j} .} As
such, $J$ may be thought of as an element of the vector space of all
closed two forms (modulo exact forms) $H^2(M,\IR)$. In fact, $J$ lies in
a special subspace of this vector space known as the K\"ahler cone by
virtue of its relation to the metric. In particular, since the metric
measures non-negative lengths, areas and volumes, $J$ satisfies
\eqn\eVol{\int_M J \wedge J \wedge J > 0}
\eqn\eArea{\int_S J \wedge J > 0}
\eqn\eLength{\int_C J > 0.}
where $S$ and $C$ are nontrivial 4 and 2-cycles on the
manifold respectively.

\iffigs
\midinsert
$$\vbox{\centerline{\epsfxsize=3.5cm\epsfysize=5cm\epsfbox{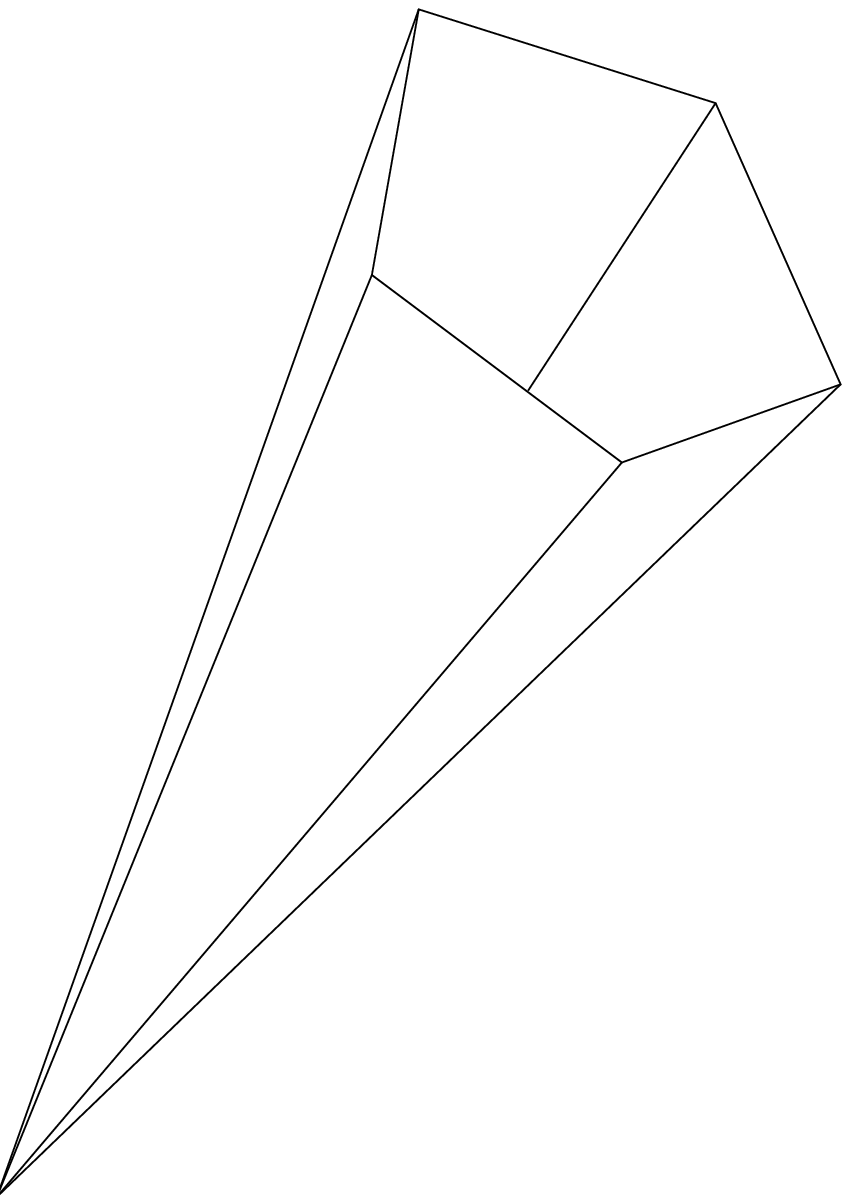}}
\centerline{Figure 1a. Schematic diagram of K\"ahler cone.}}$$
\endinsert
\fi

The space of all $J$ in $H^2(M,\IR)$ that satisfy these requirements has
a cone structure because if $J$ satisfies these conditions, so does the
positive ray generated by $J$ -- hence the name K\"ahler cone.  In figure
1a we schematically show a K\"ahler cone. A well known aspect of string
theory is that it instructs us to combine the K\"ahler form $J$ with the
antisymmetric tensor field $B$ into the complexified K\"ahler class $K =
B + iJ$.  The physical model is invariant under integral shifts of $B$
(more precisely, shifts of $B$ by elements of $H^2(M,\IZ)$) which
motivates changing variables to
\eqn\ew{ w_l = e^{2 \pi i (B_l + i J_l)} } where $(B_l, J_l)$ are
coefficients in the expansion of $B$ and $J$ with respect to an integral
basis of $H^2(M,\IZ)$. These new variables have the invariance under
integral shifts built in.

\iffigs
\midinsert
$$\vbox{\centerline{\epsfxsize=3.5cm\epsfysize=4cm\epsfbox{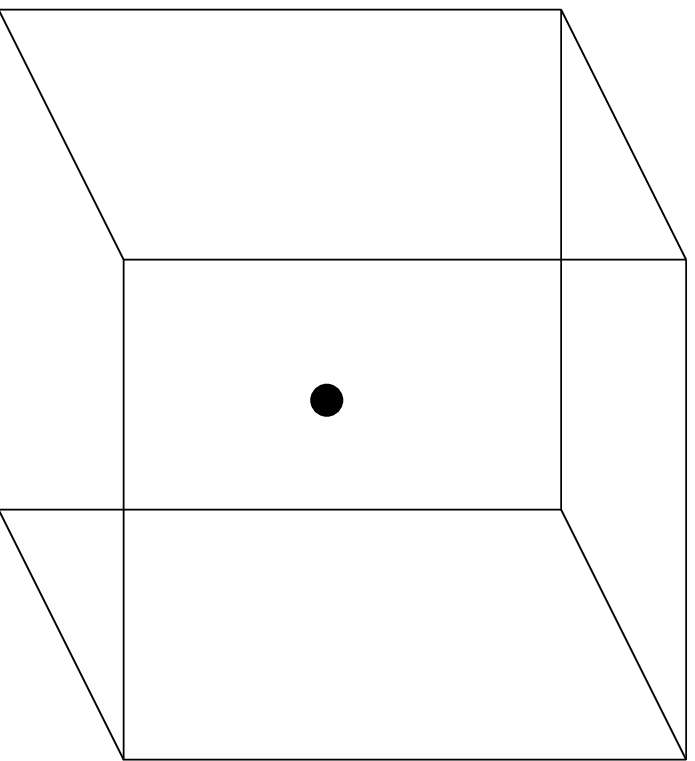}}
\centerline{Figure 1b. Complexified Kahler cone.}}$$
\endinsert
\fi

The imaginary part of $K$ satisfies the conditions on $J$ just discussed
and hence the K\"ahler cone of figure 1a becomes the bounded domain of
$H^2(M, \IC)$ in the $w$ variables as depicted in figure 1b.  We note
that the boundary of this region denotes those places in the parameter
space where the K\"ahler form $J$ degenerates in the sense that some of
the positivity requirements are violated.

We will return to this picture shortly, but first we turn to a discussion
of the typical form of complex structure moduli spaces.

\vglue0.6cm
\leftline{\tenit 2.2. Complex Structure Moduli Space}
\vglue0.4cm

 Mathematically, the choice of complex structure on a manifold is the
choice of local complex coordinates in sets of coordinate patches and
holomorphic transition functions between patches. \footnote{These choices are
equivalent if they differ by a biholomorphic change of variables.}
Physically, just as the choice of K\"ahler class determines the metric or
``size'' of the Calabi-Yau space, the choice of complex structure
determines its ``shape''. For instance, in the simplest case of a
Calabi-Yau manifold, the one-complex dimensional torus, the K\"ahler
structure fixes the overall volume while the complex structure fixes
$\tau$, the angle between the two cycles.

\iffigs
\midinsert
$$\vbox{\centerline{\epsfxsize=6cm\epsfysize=4cm\epsfbox{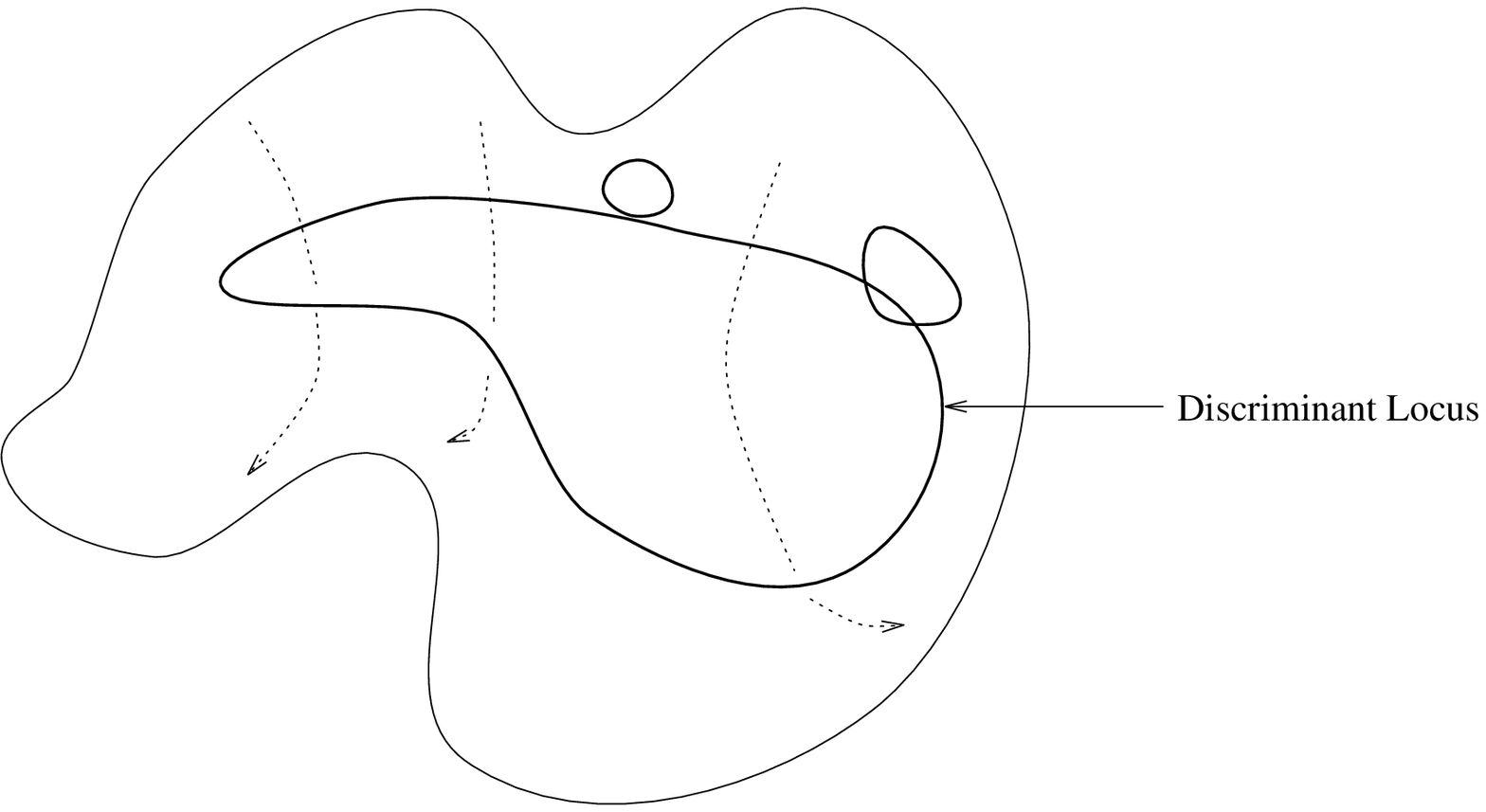}}
\centerline{Figure 2. Complex structure moduli space showing discriminant
locus.}}$$
\endinsert
\fi

The typical form of Calabi-Yau spaces we shall consider is given by the
vanishing locus of homogeneous polynomial equations in (products of)
projective spaces.  For ease of discussion, consider the case of a single
equation $P = 0$ with
\eqn\eP{P = \sum a_{i_1...i_n} z_1^{i_1}...z_n^{i_n}.} It is known that
by varying the coefficients in such equations we vary the choice of
complex structure on the the Calabi-Yau space. There is one constraint on
the choice of coefficients $a$ which we must satisfy: they must be chosen
so that $P$ and its partial derivatives do not have a common zero in the
defining projective space. If they did have such a common zero, the
choice of complex structure would be singular (non-transverse). It is
straightforward to see that this places one complex constraint on the
choice of coefficients and hence we can think of the complex structure
moduli space as the space of all $a$'s (modulo those which are equivalent
via coordinate transformations on the z's) less this one complex
constraint. Schematically, this space can be illustrated as in figure 2
where the``bad" choice of $a$'s is correctly denoted as the {\it
discriminant locus} \footnote{Recently, it has been shown that at least
some and possibly all points on the discriminant locus correspond to well
defined string theories with the singularities in the conformal field
theory description associated with the appearance of new massless degrees
of freedom
\ref\rStrom{A. Strominger, {\it Massless Black Holes and Conifolds in String
Theory}, hep-th/9504090}
\ref\rGMS{B. R. Greene, D. R. Morrison, and A. Strominger, {\it Black Hole
Condensation and the Unification of Sting Vacua}, hep-th/9504145}\
. We will not discuss this issue here.}.

\vglue0.6cm
\leftline{\tenit 2.3. The Problem}
\vglue0.4cm

The puzzle for mirror symmetry can now be described directly. Mirror
symmetry tells us that figure 1b and figure 2 are isomorphic if the
former is for $M$ and the latter for its mirror $W$. However, manifestly
they are not. This is not a product of our schematic drawings as there
are genuine qualitative distinctions. Most prominently, note that the
locus of what {\it appear} to be badly behaved theories is real
codimension one in the K\"ahler parameter space, occurring on the walls
of the domain where the classical K\"ahler form degenerates. On the
contrary, the locus of badly behaved theories in the complex structure
moduli space, as just discussed, is real codimension two (complex
codimension one). What is going on? The answer to this question was found
in
\rWittenphases\ and \rAGM\ and implies that:
\vskip.2in {$\bullet$ 1: Figure 1b for $M$ is only a {\it subset} of
figure 2 for $W$.  To be isomorphic to figure 2 of $W$, it must be
augmented by numerous other regions, of a similar structure, all adjoined
along common walls. This yields the {\it enlarged K\"ahler moduli space}
of $M$.}
\vskip.1in {$\bullet$ 2: Some of these additional regions are
interpretable as the complexified K\"ahler moduli space of {\it flops} of
$M$ along rational curves. These are relatively subtle transformations of
one Calabi-Yau manifold to yield other (birationally equivalent) possibly
topologically distinct Calabi-Yau's.}
\vskip.1in {$\bullet$ 3: Other regions may not have a direct sigma model
interpretation, but rather are the parameter spaces for Landau-Ginzburg
theory, Calabi-Yau orbifolds, and various relatively unfamiliar hybrid
combination conformal theories.}
\vskip.1in {$\bullet$ 4: Whereas classical reasoning suggests that
theories whose complexified K\"ahler class lies on the wall of a domain
such as that in figure 1b are ill defined, quantum reasoning shows that
the generic point on such a wall corresponds to a perfectly well behaved
theory. Thus, physics changes {\it smoothly} if the parameters of a model
change in a generic manner from one region to another by crossing through
such a wall. As some such regions correspond to sigma models on
topologically distinct target spaces, this last point establishes the
first concrete example of physically allowed spacetime topology change.}

We can pictorially summarize this discussion as we do in figure 3. Here
we see that the abstract conformal field theory moduli space is
geometrically interpretable in terms of the complex structure and
enlarged K\"ahler structure parameter spaces associated to $M$ or to the
enlarged K\"ahler structure and complex structure parameter spaces
associated to the mirror $W$. This picture allows us to elaborate on
point 4 above. Let's imagine following a path in the enlarged K\"ahler
moduli space of $M$ that passes through a wall between two sigma model
regions. As mentioned, at first sight it appears that the corresponding
sigma model is ill defined when the parameters are those associated with
the wall since the K\"ahler form has degenerated. Examining the quantum
theory directly, we see that the question as to whether or not it is well
behaved is hard to answer since reliable perturbation theory requires that
distances be suitably large (recall that the expansion parameter is
$\alpha' \over R^2$) while certain lengths shrink to zero on the wall.
Thus, probing the physical meaning of passing through a wall by direct
analysis does not yield a definitive conclusion. Now, though, mirror
symmetry comes to the rescue. We can rephrase the journey taken in the
K\"ahler moduli space of $M$ as a journey in the complex structure moduli
space of $W$. Furthermore, nothing in our discussion has dealt with the
complex structure of $M$, so we can choose it at our convenience (and
keep it fixed throughout the analysis). We choose it so that the
corresponding mirror K\"ahler structure in the enlarged K\"ahler moduli
space of $W$ is far from any walls and hence deeply in the perturbative
realm. In this way, the {\it strong} coupling question of analyzing the
physics of passing through a wall in the $M$ description has been
translated into a {\it weak} coupling question on $W$. Now, a journey
through the complex structure moduli space of $W$, keeping the K\"ahler
structure fixed at a value ensuring perturbative reliability, yields
perfectly well behaved physics so long as we stay away from the
discriminant locus. As the latter is real codimension two, this is simple
to do: the generic path misses the bad theories. Since we trust this
analysis, the same must be true for the physically isomorphic mirror path
on $M$ and hence physics is smooth as we pass through a generic point on
a K\"ahler wall.

\iffigs
\topinsert
$$\vbox{\centerline{\epsfxsize=8cm\epsfysize = 8cm\epsfbox{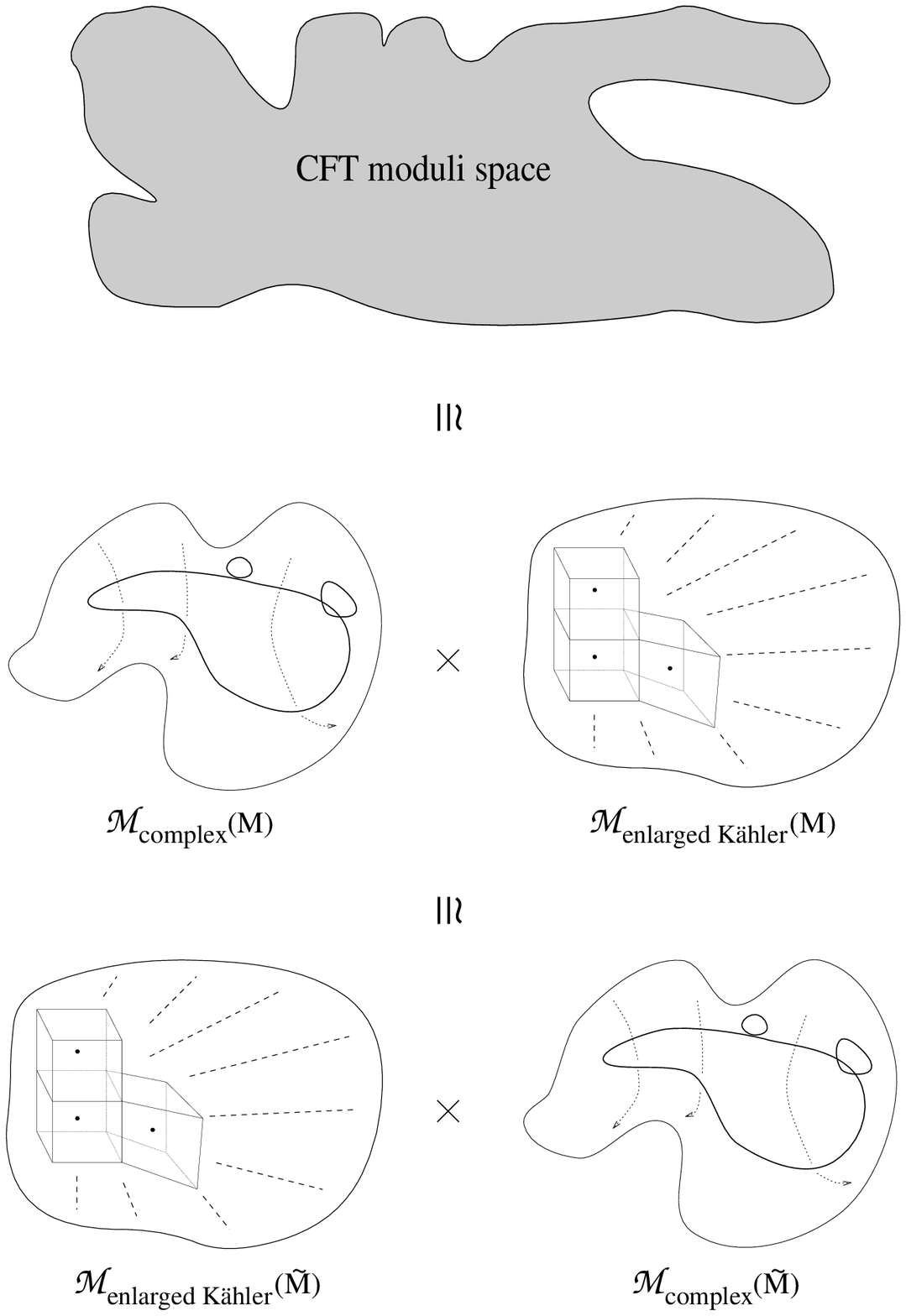}}
\centerline{Figure 3. Schematic drawing of the moduli spaces for $M$ and
it's mirror $W$}}$$
\endinsert
\fi

\vglue0.6cm
\leftline{\tenbf 3. Constructing the Phase Diagram}
\vglue0.4cm
The works of \rWittenphases\ and of \rAGM\ present two different methods
for constructing the phase diagram for any given example. Each has its
virtues so lets briefly review them both.

\vglue0.6cm
\leftline{\tenit 3.1. Toric Geometry: Symplectic Quotients}
\vglue0.4cm

In \rWittenphases, Witten starts with a gauged linear sigma model in
two dimensions and studies its vacuum structure for various values of
coefficients of Fayet-Illiopoulos D-terms. To be concrete, consider the
case of the quintic hypersurface which corresponds to gauge group $U(1)$.
The field content of this theory consists of six chiral superfields
$P,\Phi_1,...,\Phi_5$ with charges $(-5,1,1,1,1,1)$. From these fields we can
build a gauge invariant superpotential $PG(\Phi_1,...,\Phi_5)$ where
$G$ is a transverse quintic polynomial in the $\Phi's$. The detailed form of
the Lagrangian can be found in \rWittenphases; for our purpose we shall only
be concerned with the bosonic potential which takes the form
\eqn\eBosonic{U = |G|^2 + |p|^2(\sum |{{\partial G \over \partial
\phi_i}}|^2) + (\sum |\phi_i|^2 - 5|p|^2 - r)^2} where $r$ is the
coefficient of the D-term and we have dropped terms that will not play a
direct role in our discussion.

We now study the vacuum structure for various choices of $r$. In fact,
qualitatively there are only two distinct cases: $r$ positive and $r$
negative. We consider each in turn. When $r$ is positive we learn that
not all of the $\phi_i$ can vanish simultaneously. By the transversality
of $G$ we then learn that $p$ must vanish. Thus, the vacuum is given by
$G = 0$ inside of $\sum |\phi_i|^2 = r$.  In fact, since we are dealing
with a $U(1)$ gauge theory we have the further identification
\eqn\ephase{(\phi_1,...,\phi_5) \equiv
(e^{i\theta}\phi_1,...,e^{i\theta}\phi_5)} Now it is straightforward to
see that the $\phi_i$ span $\IC \IP^4$. The usual identification
\eqn\ephases{(\phi_1,...,\phi_5) \equiv (\lambda \phi_1,...,\lambda
\phi_5)} for arbitrary nonzero complex parameter $\lambda$ is here
realized in two steps:  we pick out one point in the equivalence class by
first imposing $\sum |\phi_i|^2 = r$, which determines the magnitude of
$\lambda$ and then by imposing \ephase\ which eliminates the remaining
phase ambiguity.  Thus, for $r$ positive the vacuum state is the
vanishing of the quintic $G$ in $\IC \IP^4$, the most well known
Calabi-Yau manifold.

For $r$ negative we proceed in a like fashion. We immediately find that
$p$ can not be zero and that all of the $\phi_i$ must vanish. Using the
gauge freedom to fix the phase of $p$ we thus learn that we have massless
fluctuating modes about a unique vacuum state -- in other words a
Landau-Ginzburg model. Thus we see that this linear sigma model
effectively has two phases: a Calabi-Yau sigma model phase and a
Landau-Ginzburg phase. Three points of clarification are called for.

{$\bullet$ 1:} The parameter $r$ is actually the real part of a complex
paramter $t$ (whose imaginary part is a theta angle). Thus each phase is
actually two real dimensions meeting along a common one dimensional locus.
After suitable compactification, we can think of these two regions as the
northern and southern hemispheres meeting along the equator. As argued by
Witten, the conformal field theory is well behaved so long as $\theta$ is
not zero, and hence there is no obstruction to passing from one phase to
another.
\vskip.1in {$\bullet$ 2:} It is worthwhile to think about the D-term part
of the vacuum analysis just described as separate from the full bosonic
potential discussion. For $r$ positive the space is
$(\phi_1,...,\phi_5;p)$ with not all $\phi_i$ zero simultaneously subject
to the (gauge) equivalence relation
\eqn\eequiv{(\phi_1,...,\phi_5;p) \equiv (\lambda \phi_1,...,\lambda
\phi_5;\lambda^{-5}p).} This space is recognized as the line bundle
${\cal O}(-5)$ over $\IC \IP^4$.  For $r$ negative, since $p$ is nonzero,
we can solve the equivalence relation by setting $p$ to one, modulo a
left over $\BZ_5$ ambiguity.  Thus, the space of fields is $\IC^5/\BZ_5$.
Mathematically, the line bundle ${\cal O}(-5)$ over $\IC \IP^4$ and
$\IC^5/\BZ_5$ are very closely related: they are birationally equivalent.
This simply means that they agree except on lower dimensional subspaces.
One can see this by recalling that the blow up of the origin in
$\IC^5/\BZ_5$ is the line bundle ${\cal O}(-5)$ over $\IC \IP^4$. Thus,
away from this exceptional set, they agree. The interpolation from one
phase to another is thus seen as a particular kind of birational
transformation on the embedding space. In fact, setting the D-term $(\sum
|\phi_i|^2 - 5|p|^2 - r)$ to zero is, mathematically speaking, known as
setting a {\it moment map} to zero. Combining this with the $U(1)$ gauge
equivalence is known as performing a symplectic quotient. The resulting
ambient spaces are {\it toric varieties} realized via symplectic
quotients.

{$\bullet$ 3:} For ease of discussion we have focused on the case in
which the K\"ahler moduli space is one dimensional, i.e. a single U(1)
gauge group. More generally, this number can be larger (or smaller, as we
shall discuss shortly) yielding a much richer phase structure.  For
instance, if the enlarged K\"ahler moduli space is $n$ dimensional, we
will have a $U(1)^n$ gauge group and parameters $(r_1,...,r_n)$
multiplying the respective Fayet-Illiopoulos D-terms.  In fact, in many
examples there are different phase regions corresponding to nonlinear
sigma models on topologically distinct Calabi-Yau target spaces. The
ability to freely move from one phase region to another in such examples
gives rise to physically smooth topology changing processes.

\vglue0.6cm
\leftline{\tenit 3.2. Toric geometry: Holomorphic quotients}
\vglue0.4cm

The symplectic approach to building the embedding space of our models is
directly transcribable, as we have seen, into the language of Lagrangian
dynamics. The equivalent holomorphic approach is somewhat more removed from
the physical description, but is often more powerful for understanding the
phase structure.

The mathematical ideas are quite simple. Consider $\ICP^n$ realized as
${\IC^{n+1} - (0,...,0)} \over \IC^*$. This is the simplest example of a
toric variety realized as a holomorphic quotient: we start with
$\IC^{n+1}$, remove some set of lower dimension (in this case just the
origin) and holomorphically divide by $\IC^*$. This immediately yields a
natural generalization: start with $\IC^{n+1}$, remove some set that we
shall call $F_{\Delta}$ and holomorphically divide by $(\IC^*)^m$. Two
such toric varieties which differ only by the set $F_{\Delta}$ which is
removed are birationally equivalent (as clearly they agree everywhere
except on a set of lower measure).

Lets consider the example discussed in the previous subsection, now in the
holomorphic language. Begin with $\IC^6$ with six complex ``coordinates"
$(\phi_1,...,\phi_5;p)$. On this space define a $\IC^*$ action
\eqn\eactionstar{(\phi_1,...,\phi_5;p) \rightarrow
 (\lambda \phi_1,...,\lambda \phi_5;\lambda^{-5} p).} Let $\F_{\Delta_1}$
be $(0,...,0;p)$ and $\F_{\Delta_2}$ be $(\phi_1,...,\phi_5;0)$.  The
first choice yields ${\cal O}(-5)$ over $\IC \IP^4$ and the second yields
$\IC^5/\BZ_5$.

The important point for our purpose here is that the holomorphic approach
gives a simple combinatorial method for working out the phase diagram of
any example as we now briefly describe. The different phase regions of a
given example constitute a particular class of birational transformations
on any ``seed'' member of the phase structure. In the holomorphic
approach, these different birational models are specified by different
choices for the set $\F_{\Delta}$. The set of allowed choices for the
$F_{\Delta}$ are in one to one correspondence with the set of
triangulations of a particular point set in $\IR^n$ for an n-dimensional
toric variety.  We will not discuss this point set in detail here, as it
is fully discussed in \rAGM\ and in \rAG. Our point is simply to
emphasize that whereas in the symplectic approach one needs to find
regions in the parameter space of $(r_1,...,r_n)$ corresponding to the
same vacuum state of the linear sigma model, in the holomorphic approach
one only needs to find the set of distinct triangulations of a point set
in $\IR^n$.

Just to give a feel for how this goes in a simple example, lets again return
to the case of the quintic. Consider the point set
$$\offinterlineskip \eqalign{ \alpha_1&=(1,0,0,0,1)\cr
	\alpha_2&=(0,1,0,0,1)\cr
	\alpha_3&=(0,0,1,0,1)\cr
	\alpha_4&=(0,0,0,1,1)\cr
	\alpha_5&=(-1,-1,-1,-1,1)\cr
	\alpha_6&=(0,0,0,0,1)\cr}$$
in $\IR^5$.
(To understand  why this is the relevant point set to consider, consult
\rAGM.)
 The algorithm for finding the possible $F_{\Delta}$ requires that
we find the possible triangulations of this point set. There are two:
The first corresponds to the Calabi-Yau phase and is given by the union of
5 cones, each of which contains 4 points chosen from the set $\{\alpha_1,
\alpha_2, \alpha_3, \alpha_4, \alpha_5\}$ and $\alpha_6$, and the
second gives the Landau-Ginzburg phase with only one cone $\{\alpha_1,
\alpha_2, \alpha_3, \alpha_4, \alpha_5\}$.

A lower dimensional version of this, as given in the figure, may help
with visualization. In this figure, the central point on the top face
denotes $\alpha_6$ and we see that the two triangulations differ by
whether or not this point is included. The lowest point represents the
origin of the space.

\iffigs
\midinsert
$$\vbox{\centerline{\epsfxsize=8cm\epsfbox{strings4.ps}}
\centerline{Figure 4. The Calabi-Yau and Landau-Ginzburg phase for the
quintic.}}$$
\endinsert
\fi

Now, to each triangulation,
as discussed in \rAGM, we find the point set $\F_{\Delta}$ according to

\eqn\efset{\bigcap_{\sigma\in\Delta} \Bigl\{ x\in\IC^6;
  \prod_{{\alpha_i\not\in\sigma}}\!\!x_i=0 \Bigr\}}
where $\sigma$ denotes a cone in the fan $\Delta$.

We see that this formula, applied to the two triangulations, does in fact
yield the two point sets discussed above.

One feature that we have not mentioned is how, in examples with more than
two phase regions in a higher dimensional setting, we glue the various
phase regions together. There is also a simple combinatorial way of doing
this as discussed in \rAGM, but for lack of time we shall not go into
that here.

\vglue0.6cm
\leftline{\tenit 3.3. Generalization:}
\vglue0.4cm

In our discussion so far we have implicitly assumed that we start with
some Calabi-Yau manifold embedded in some toric variety such as a
weighted projective space or products thereof. We can then fill out the
full phase diagram, of which this Calabi-Yau description is but one
member, by using either the symplectic or holomorphic formalism described
above. In the former, we realize the toric embedding space via a gauged
linear sigma model, and then determine the form of all possible vacuum
states by varying the Fayet-Illiopoulos D-terms. In the latter we realize
the embedding space in the formalism of toric geometry and then determine
all K\"ahler birational transformations via a combinatorial procedure.
Given such a phase diagram, no one region is any more special than any
other and hence one might suspect that one need not start out from a
Calabi-Yau manifold in a toric variety.  Rather, in the symplectic
formalism we may start with any linear sigma model with the requisite
nonanomalous R-symmetries that flow in the infrared to the $U(1)$
currents in the N = 2 superconformal algebra. In the holomorphic
formalism, we can start with any toric point set data with suitably mild
singularities that, again, gives rise in the corresponding physical model
to the necessary nonanomalous R-symmetries. Technically, such toric data
corresponds to that for {\it reflexive Gorenstein cones} \ref\rBatyrev{V.
Batyrev, {\it Dual Polyhedra and Mirror Symmetry for Calabi-Yau
Hypersurfaces in Toric Varieties}, J. Alg. Geom. {\bf 3} (1994)
493-535.}.
\footnote{For more details in language geared to physicists see \rAG.}
Given such data, in either formalism, one can then build the phase
diagram, in the manner outlined above. An important fact is that, for a
general example, nothing guarantees that the phase diagram will contain a
Calabi-Yau sigma model region or even a Landau-Ginzburg region.  The fact
that there may not be a Landau-Ginzburg region was explicitly shown in
some of the examples in \rWittenphases\ in which, for instance, the
closest cousin to a Landau-Ginzburg theory is a gauged Landau-Ginzburg
theory.  Below we shall see some examples that lack a Calabi-Yau sigma
model region.

Our program for building phase diagrams, using the holomorphic
formalism, is thus clear:

{$\bullet$ 1. Select a point set corresponding to a reflexive Gorenstein cone.}

{$\bullet$ 2. Find all possible (regular) triangulations.}

{$\bullet$ 3. Using these triangulations, build the phase diagram and
   establish the physical identity of each phase.}

We now present the results of such a program.

\vglue0.6cm
\leftline{\tenbf 4. Generic Examples of Phase Diagrams}
\vglue0.4cm

In the first class of examples, we start with toric data used to describe
(complete intersection) Calabi-Yau manifolds in (products of weighted)
projective space(s).  Thus, in these examples, we are guaranteed to have
a Calabi-Yau sigma model region. Below we present a number of examples,
of increasing K\"ahler moduli space dimension, to illustrate the type of
phase diagrams which arise.

We start first with a degree-18 hypersurface in the weighted projective
space $\IP^{4}_{(6,6,3,2,1)}$. It has a $h^{1,1}$ = 7 and $h^{2,1}$ =
79. As discussed in \rAGM, only a 5-dimensional subspace of the
$(1,1)$-forms on the manifold is realized torically. The point set
describing the toric variety is

$$\offinterlineskip \eqalign{ \alpha_1&=(-3,-3,-1,-1,1)\cr
	\alpha_2&=(-2,-2,-1,0,1)\cr
	\alpha_3&=(-4,-4,-2,-1,1)\cr
	\alpha_4&=(-1,-1,0,0,1)\cr
	\alpha_5&=(1,0,0,0,1)\cr
	\alpha_6&=(0,1,0,0,1)\cr
	\alpha_7&=(0,0,1,0,1)\cr
	\alpha_8&=(0,0,0,1,1)\cr
	\alpha_9&=(-6,-6,-3,-2,1)\cr}$$

The 100 triangulations of this set \rAGM\ contains 5 Calabi-Yau
resolutions which can be specified by looking at the face spanned by
$\{\alpha_7, \alpha_8,
\alpha_9\}$, depicted in figure 5.

\iffigs
\midinsert
$$\vbox{\centerline{\epsfysize=7cm\epsfbox{strings5.ps}}
\centerline{Figure 5. The 5 smooth resolutions for $\wgt4 6 6 3 2 1$}}$$
\endinsert
\fi

Besides the 5 Calabi-Yau regions, there is the obvious Landau-Ginzburg
region comprised of the single simplex $\{\alpha_5, \alpha_6, \alpha_7,
\alpha_8, \alpha_9\}$ having volume 18. Then there are 27 Calabi-Yau
orbifolds obtained by special star {\it unsubdivisions} of the Calabi-Yau
phases we had, that is, by removing an interior point from a phase that
gives us a new triangulation with all simplices still having the origin
$(0,0,0,0,1)$. This gives rise to $\IZ_{2}, \IZ_{3}, \IZ_{2} \times
\IZ_{2},
\IZ_{4} \hbox{ and } \IZ_{6}$ quotient singularities. The remaining 67
triangulations correspond to Landau-Ginzburg {\it hybrid models}.

	We may build the {\it secondary fan} which describes the moduli
space of the phases using a technique described in \rAGM. In figure 6 we
depict the connectivity of the phases in the above example by plotting
vertices chosen from the interior of each separate phase.

\iffigs
\midinsert
$$\vbox{\centerline{\epsfysize=6cm\epsfbox{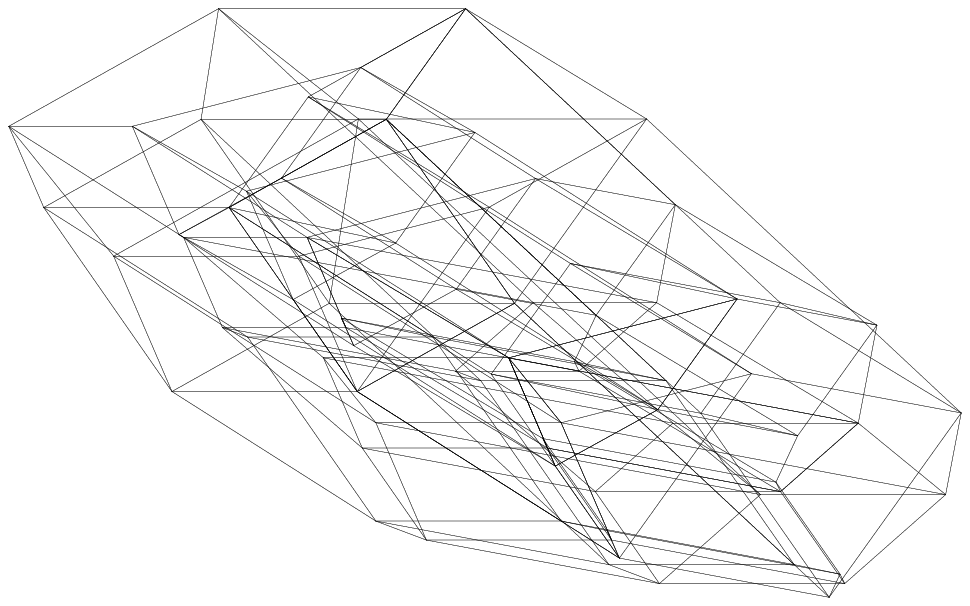}}
\centerline{Figure 6. The connectivity of the 100 phases of $\wgt4 6 6 3
2 1$}}$$
\endinsert
\fi

\vskip.3in

	Now we'll turn to examples involving CICYs (Calabi-Yau
complete-intersections) in products of ordinary projective spaces. A CICY
(more precisely a family of CICYs) is described by a configuration matrix:

\eqn\ecicyconfig{\left(\matrix n_1 \cr n_2 \cr\vdots \cr n_r \cr
\endmatrix \right|
	\left|\matrix a_{11}&a_{12}&\ldots&a_{1k}\cr
		a_{21}&a_{22}&\ldots&a_{2k}\cr
		\vdots&\vdots&\ddots&\vdots\cr
		a_{r1}&a_{r2}&\ldots&a_{rk}\cr \endmatrix\right)}

Here $n_i$ is a shorthand for $\IP^{n_i}$ and the $a_{ij}'s$ are
interpreted as degrees of $k$ polynomials which intersect transversely, the
$j^{th}$ of which is homogenous of degree $a_{ij}$ in the variables of
$\IP^{n_i}$.

	In the phases picture we may formulate the problem as follows.
Let us associate to the $i^{th}$ projective space the chiral superfields
$\phi_{il}$ where $l$ runs from $1$ to $n_i + 1$. We will be considering a
$U(1)^r$ gauged linear sigma model based on the $\phi$'s as well as $k$
other chiral superfields $p_{j}$ representing the constraints. $p_j$ will
have charge $-a_{ij}$ under the $i^{th}$ $U(1)$. Thus we may form the
gauge-invariant superpotential

\eqn\esuperpotential{W = \sum_{i=1}^{k} p_{i} G_{i}(\{\phi_{jk}\})}

where $G_{i}$ is the $i^{th}$ constraint and the bosonic potential has the
following form:

\eqn\ecicybosonic{U = \sum_{i=1}^{k} |{\partial W \over \partial p_{i}}|^2
+ \sum_{i=1}^{r} \sum_{j=1}^{n_i+1} |{\partial W \over \partial \phi_{ij}}|^2 +
\sum_{i=1}^{r} (\sum_{j=1}^{n_i+1} |\phi_{ij}|^2 - \sum_{j=1}^{k} a_{ij}
|p_{j}|^2 - r_{i})^2}

Transversality of the $G_{i}$'s imply that the conditions neccesary
for the minimization of the potential:

\eqn\eminpotential{{\partial W \over \partial p_{1}} = \ldots = {\partial W
\over \partial p_{k}} = {\partial W \over \partial \phi_{11}} = \ldots =
{\partial W \over \partial \phi_{r{n_r+1}}} = 0} {\noindent}are realized
if and only if the set of fields $p_{i}$ all vanish or the $\phi_{ij}$'s
vanish for some $i$.

Again, one could work out the phase diagram by studying the semi-classical
vacua for various choices of the $r$ parameters and grouping together
those which yield the same phase. We follow the holomorphic approach by
realizing the ambient toric variety (which here is just a product of
ordinary projective spaces) as the associated point set and then seeking
out all possible triangulations. It is worth noting two things.  First,
distinct phases in the ambient toric variety may not correspond to
identical phases in the actual physical model. From the toric view this
is simply the statement that the Calabi-Yau is of finite codimension in
the toric variety and hence may not intersect the distinct $F_{\Delta}$'s
which distinguish two phases.  Physically, the vanishing of the F-terms
may not intersect the point set distinguishing between the vanishing
locus of distinct D-terms (which differ by the $r$'s lying in different
phase regions).  Second, a given Calabi-Yau manifold can often be
realized as embedded in distinct ambient toric varieties.
 It turns out that moduli having a toric representation in one embedding may
not have a toric representation in another and vice versa. Thus, different
embeddings, although yielding topologically identical Calabi-Yau's can provide
different moduli space information.

	In the following we show a few examples of hypersurfaces and
CICYs and the total number of triangulations as well as (in the case of
CICYs) the number of smooth phases \footnote{By a smooth phase we refer
to those phases embedded in a smooth ambient toric variety}. The examples
chosen illustrate a dramatic increasing trend of the number of phases
with the value of $h^{1,1}_{toric}$.  We have actually been rather
selective in our presentation for the CICYs. The collection of results
for about 8000 configurations with $h^{1,1}_{toric}$ ranging from 1 to 8
show that one can find a large range in the number of phases for any $r$
and these ranges do intersect widely. It remains true that the maximum
number of phases (or smooth phases) found increases with
$h^{1,1}_{toric}$. It is interesting to note that the $h^{1,1}_{toric} =
6$ example shown on the next page has all of it's phases being smooth.
We'll say more about this in the next section.

	In the following we give some examples of the number of phases
found for hypersurfaces in various weighted projective spaces with 5
weights:

\midinsert
\def\tablerule{\noalign{\hrule}}
$$\vbox{\tabskip=0pt \offinterlineskip
\halign{
\strut#& \vrule#& \hfil \quad$#$\quad\hfil &\vrule#&
\hfil\quad# \quad\hfil&\vrule#&
\hfil\quad# \quad\hfil&\vrule#&
\hfil\quad# \quad\hfil&\vrule#&
\hfil\quad# \quad\hfil&\vrule#\cr
\tablerule
&height 15pt depth 10pt
	&\hbox{Manifold}
	   &&$h^{1,1}$
		&&$h^{2,1}$
	          &&$h^{1,1}_{toric}$
		    &&\hbox{Phases}
			&\cr\tablerule
&height 15pt
	&\wgt4 1 1 1 1 1
	    &&1&&101&&1
	    &&2&\cr\tablerule
&height 15pt
        &\wgt4 3 3 1 1 1
            &&4&&112&&2
            &&4&\cr\tablerule
&height 15pt
	&\wgt4 5 3 3 3 1
	   &&3&&75&&3
	   &&12&\cr\tablerule
&height 15pt
	&\wgt4 5 5 5 4 1
	    &&5&&65&&5
	    &&56&\cr\tablerule
&height 15pt
	&\wgt4 7 7 3 3 1
	    &&17&&65&&5
	    &&82&\cr\tablerule
&height 15pt
	&\wgt4 6 6 3 2 1
	    &&7&&79&&5
	    &&100&\cr\tablerule
&height 15pt
	&\wgt4 8 6 6 3 1
	    &&8&&68&&5
	    &&157&\cr\tablerule
&height 15pt
	&\wgt4 8 8 4 3 1
	   &&10&&70&&6
	    &&296&\cr\tablerule
&height 15pt
	&\wgt4 10 10 6 3 1
	   &&19&&67&&7
	    &&1709&\cr\tablerule
&height 15pt
	&\wgt4 12 12 9 2 1
	    &&16&&100&&8
	     &&4854&\cr\tablerule
&height 15pt
	&\wgt4 20 15 12 12 1
	   &&23&&47&&11
	   &&$>=$ 703488&\cr\tablerule
}}
$$
\endinsert
\break

\pageinsert
\def\tablerule{\noalign{\hrule}}
$$\vbox{\tabskip=0pt \offinterlineskip \textfont0=\sevenrm
\halign{
\strut#& \vrule#& \hfil \quad$#$\quad\hfil &\vrule#&
\hfil\quad# \quad\hfil&\vrule#&
\hfil\quad# \quad\hfil&\vrule#&
\hfil\quad# \quad\hfil&\vrule#\cr
\tablerule
&height 15pt depth 10pt
	&\hbox{Manifold}
	   &&$h^{1,1}_{toric}$
		    &&\hbox{Phases}
			&&\hbox{Smooth phases}&\cr
\tablerule
&height 10pt
	& \left(4\right|\left|5\right)
	    &&1
	    &&2&&1&\cr
&height 15pt
	&\left(\matrix 3 \cr 2 \endmatrix \right|\left|
	\matrix 2  & 2 \cr 2 & 1 \endmatrix \right)
	    &&2
	&&4&&1&\cr
&height 20pt
	&\left(\matrix 4 \cr 1 \cr 1 \endmatrix \right|\left|
	\matrix 2 & 2 & 1 \cr 0 & 0 & 2 \cr 1 & 0 & 1 \endmatrix \right)
	&&3
	&&6&&1&\cr
&height 30pt
	&\left(\matrix3 \cr 3 \cr 2 \cr 1 \endmatrix \right|\left|
	\matrix 2 & 1 & 1 & 0 & 0 & 0 \cr
	0 & 1 & 0 & 2 & 1 & 0 \cr
	0 & 0 & 1 & 0 & 1 & 1 \cr
	0 & 0 & 0 & 1 & 0 & 1 \endmatrix\right)
	&&4
	&&130&&5&\cr
&height 35pt
	&\left(\matrix 2 \cr 2 \cr 2 \cr 2 \cr 2 \endmatrix \right|\left|
	\matrix 1 & 1 & 1 & 0 & 0 & 0 & 0 \cr
	1 & 0 & 0 & 1 & 1 & 0 & 0 \cr
	0 & 1 & 0 & 1 & 0 & 1 & 0 \cr
	0 & 0 & 1 & 1 & 0 & 0 & 1 \cr
	0 & 0 & 0 & 0 & 1 & 1 & 1 \endmatrix \right)
	&&5
	&&22784&&302&\cr
&height 45pt &\left(\matrix 2 \cr 2 \cr 2 \cr 2 \cr 2 \cr 2 \endmatrix
	\right|\left| \matrix 1 & 1 & 1 & 0 & 0 & 0 & 0 & 0 & 0 \cr 1 & 0
	& 0 & 1 & 1 & 0 & 0 & 0 & 0 \cr 0 & 1 & 0 & 0 & 0 & 1 & 1 & 0 & 0
	\cr 0 & 0 & 1 & 0 & 0 & 0 & 0 & 1 & 1 \cr 0 & 0 & 0 & 1 & 0 & 1 &
	0 & 1 & 0 \cr 0 & 0 & 0 & 0 & 1 & 0 & 1 & 0 & 1 \cr \endmatrix
	\right) &&6 &&33415&&33415&\cr
&height 50pt
	&\left(\matrix 2 \cr 1 \cr 1 \cr 1 \cr 1 \cr 1 \cr
1 \endmatrix \right|\left|
	\matrix 1 & 1 & 1 & 0 & 0 \cr
	1 & 0 & 0 & 1 & 0 \cr
	0 & 1 & 0 & 1 & 0 \cr
	0 & 0 & 1 & 1 & 0 \cr
	1 & 0 & 0 & 0 & 1 \cr
	0 & 1 & 0 & 0 & 1 \cr
	0 & 0 & 1 & 0 & 1 \cr \endmatrix \right)
	&&7
	&&22690&&6168&\cr
&height 60pt
	&\left(\matrix 2 \cr 1 \cr 1 \cr 1 \cr 1 \cr 1 \cr 1 \cr
1 \endmatrix \right|\left|
	\matrix 1 & 1 & 1 & 0 & 0 & 0 \cr
	1 & 0 & 0 & 1 & 0 & 0 \cr
	1 & 0 & 0 & 0 & 1 & 0 \cr
	0 & 1 & 0 & 1 & 0 & 0 \cr
	0 & 0 & 1 & 0 & 1 & 0 \cr
	0 & 0 & 0 & 1 & 0 & 1 \cr
	0 & 0 & 0 & 0 & 1 & 1 \cr
	0 & 0 & 0 & 0 & 0 & 2 \cr \endmatrix \right)
	&&8
	&&39772&&6484&\cr
&height 5pt
	&&&&&&&&\cr
\tablerule}}
$$
\endinsert

\leftline{\tenbf 5. Special Examples of Phase Diagrams}
\vglue0.4cm

Whereas in the previous examples, by our starting point we were guaranteed
to have a Calabi-Yau sigma model region, here we note some examples which
do not share this property. We will also see some some examples in which
there is no Landau-Ginzburg region, gauged or ungauged.

\leftline{\tenbf Example 1}

Consider the point set
$${\offinterlineskip \eqalign{\alpha_1&=(4,-3,0,0,0,0,0,0,0)\cr
   \alpha_2&=(0,1,0,0,0,0,0,0,0)\cr
   \alpha_3&=(0,0,1,0,0,0,0,0,0)\cr
   \alpha_4&=(0,0,0,1,0,0,0,0,0)\cr
   \alpha_5&=(0,0,0,0,1,0,0,0,0)\cr
   \alpha_6&=(0,0,0,0,0,1,0,0,0)\cr
   \alpha_7&=(0,0,0,0,0,0,1,0,0)\cr
   \alpha_8&=(-4,2,-2,-1,-1,-2,-1,4,-2)\cr
   \alpha_9&=(0,0,0,0,0,0,0,0,1)\cr
   \alpha_{10}&=(3,-2,0,0,0,0,0,0,0)\cr
   \alpha_{11}&=(2,-1,0,0,0,0,0,0,0)\cr
   \alpha_{12}&=(1,0,0,0,0,0,0,0,0)\cr}}$$
\break
used in \rAG\ to describe the toric variety $\IC^9/(\IZ_4 \times \IZ_4)$
whose $h^{1,1} = 1$.  The 8 triangulations that occur here are easily
described by noticing that $\alpha_1, \alpha_2$ form a line with
$\alpha_{10}, \alpha_{11},
\alpha_{12}$ being points on it going from $\alpha_1$ to $\alpha_2$. A
triangulation of the cone corresponds to a choice of points among
$\alpha_{10}, \alpha_{11}, \alpha_{12}$ to include in the fan. If we
don't include any of the 3 points, we obtain the original space
$\IC^9/(\IZ_4 \times \IZ_4)$, physically a LG model. Including all 3
points results in hybrid LG model based on a $\IP_1$ vacuum. Other
triangulations correspond to one of the above described phases.  Notice
that there is {\it not} a smooth Calabi-Yau sigma model phase for this
theory. This means that in no phase can we properly interpret the (single)
marginal operator whose charges qualify it for the role of a K\"ahler
modulus as a geometrical radial mode. There is thus no ``large radius
limit'' for this example -- no limit in which classical geometrical
reasoning can be used to describe the physical model.

\leftline{\tenbf Example 2}

In the last example we described a case where a potential K\"ahler form
was present but no Calabi-Yau phase was to be found. Now we give an
example where there is no K\"ahler form and show how we can interpret
this within the phases picture and indicate its implication for mirror
symmetry.

We focus on the Z-manifold and its putative mirror. Recall that the
Z-manifold has $h^{1,1} = 36$ and $h^{2,1} = 0$, the latter implying that
it is {\it rigid}.  This seems to provide a puzzle for mirror symmetry as
the putative mirror would have $h^{1,1} = 0$ and hence would not be K\"ahler.
Let analyze this in the context of the phases picture.

As discussed in \rAG\  given the toric
data for the Z-manifold, we can apply the mirror construction of
\ref\rGP{B. R. Greene and M. R. Plesser, {\it Duality in Calabi-Yau Moduli
Space}, Nucl. Phys. {\bf B338} (1990) 15-37}\ (as implemented by
\rBatyrev) to yield the toric data for the mirror.  This yields the
pointset:

$${\offinterlineskip \eqalign{\alpha_1&=(3,0,0,1,1,1,-1,-1,-3)\cr
   \alpha_2&=(0,1,0,0,0,0,0,0,0)\cr
   \alpha_3&=(0,0,1,0,0,0,0,0,0)\cr
   \alpha_4&=(0,0,0,1,0,0,0,0,0)\cr
   \alpha_5&=(0,0,0,0,1,0,0,0,0)\cr
   \alpha_6&=(0,0,0,0,0,1,0,0,0)\cr
   \alpha_7&=(0,0,0,0,0,0,1,0,0)\cr
   \alpha_8&=(0,0,0,0,0,0,0,1,0)\cr
   \alpha_9&=(0,-1,-1,-2,-2,-2,0,0,3)\cr}}$$

Given these points, we can now find all triangulations to yield the phase
diagram. However, the above simplex of 9 points in 9 dimensions does not
contain any interior points, thus there is only one triangulation, namely
the simplex itself with volume 9. Mathematically, this corresponds to the
toric variety $\IC^9/(\IZ_3 \times \IZ_3)$. Physically, this is a
Landau-Ginzburg theory.  Thus, this enlarged K\"ahler moduli space
consists of one-point, hence one phase, which by direct inspection in
interpretable as a Landau-Ginzburg model.  Quite consistently, there is
no Calabi-Yau sigma model phase, as in the last example. Thus, there is
no puzzle by the absence of a K\"ahler form since there is no geometrical
phase in which it would be needed. Hence, in the generic Calabi-Yau
example, the enlarged K\"ahler moduli spaces for both the manifold and
its mirror contain smooth Calabi-Yau sigma model regions.  Manifolds
associated with these regions are called ``mirror manifolds''.  Mirror
symmetry, though, is more general --- and makes a one to one association
between all points in the original moduli space with those in the mirror
moduli space.  In the degenerate case of a rigid Calabi-Yau manifold, the
enlarged K\"ahler moduli space for its mirror is similarly degenerate: it
lacks a Calabi-Yau sigma model region, precisely in keeping with the need
to have $h^{1,1} = 0$. Thus, there is still a one to one association of
theories, but none of these involve two Calabi-Yau manifolds.

In the context of the phases picture, therefore, the puzzle of identifying
the mirror to a rigid Calabi-Yau manifold evaporates. This in no way
precludes the possibility of there being some other geometrical
interpretation, as has been proposed in
\ref\rCDP{P. Candelas, E. Derrick and L. Parkes, {\it Generalized
Calabi-Yau Manifolds and the Mirror of a Rigid Manifold}, Nucl. Phys.
{\bf B407} (1993) 115.} \ref\rSchimmrigk{R. Schimmrigk, {\it Critical
Superstring Vacua from Noncritical Manifolds: A Novel Framework for String
Compactification}, Phys. Rev. Lett. {\bf 70} (1993) 3688.}
\ref\rSethi{S. Sethi, {\it Supermanifolds, Rigid Manifolds and Mirror
Symmetry}, Harvard 1994 preprint HUTP-94-A002, hep-th/9404186}. It
simply shows that such an interpretation is not necessary with our more
robust understanding of conformal field theory moduli space.

\leftline{\tenbf Example 3}

Now let's look at a $h^{1,1} = 1$ example where we have no
Landau-Ginzburg phase, and which also illustrates a rather special $\IZ_2$
symmetry relating the 2 phases.  We consider here a complete intersection
of 4 quadrics in $\ICP^7$ giving rise to a Calabi-Yau manifold. The point
set here is given by

$${\offinterlineskip \eqalign{\alpha_1&=(1,0,0,0,0,0,0,1,0,0,0)\cr
	\alpha_2&=(0,1,0,0,0,0,0,1,0,0,0)\cr
	\alpha_3&=(0,0,1,0,0,0,0,0,1,0,0)\cr
	\alpha_4&=(0,0,0,1,0,0,0,0,1,0,0)\cr
	\alpha_5&=(0,0,0,0,1,0,0,0,0,1,0)\cr
	\alpha_6&=(0,0,0,0,0,1,0,0,0,1,0)\cr
	\alpha_7&=(0,0,0,0,0,0,1,0,0,0,1)\cr
	\alpha_8&=(-1,-1,-1,-1,-1,-1,-1,0,0,0,1)\cr
	\alpha_9&=(0,0,0,0,0,0,0,1,0,0,0)\cr
	\alpha_{10}&=(0,0,0,0,0,0,0,0,1,0,0)\cr
	\alpha_{11}&=(0,0,0,0,0,0,0,0,0,1,0)\cr
	\alpha_{12}&=(0,0,0,0,0,0,0,0,0,0,1)\cr}}$$

	There are only two phases in this theory. One is the obvious
Calabi-Yau phase given by the triangulation consisting of 8 simplices,
where each simplex consists of $\alpha_{10}, \alpha_{11}, \alpha_{12}$
and 8 points chosen from $\alpha_1 \ldots \alpha_9$. Here the
corresponding $F_{\Delta}$ is $(0,0,0,0,0,0,0) \times \IC^4$. The other
phase corresponds to the symmetric operation of removing $\IC^7 \times
(0,0,0,0)$ and this results in a Landau-Ginzburg theory fibered over
$\IP^3$.  The triangulation for this phase, as might be expected,
consists of 4 simplices of volume 2, each given by
$\alpha_1\ldots\alpha_8$ and a choice of 3 points amongst
$\alpha_9\ldots\alpha_{12}$. The $\IZ_2$ fibration over $\IP_3$ is easily
understood from the charges of p-fields: When we divide the original
$S^8$ target space $|p_1|^2 + |p_2|^2 + |p_3|^2 + |p_4|^2 = r$ by the
$U(1)$ to determine a representative in $\IP_3$, the s-fields are
identified up to a sign change due to the charges of the $p$'s. It's
instructive to see how this comes about in the toric language. Consider
any 2 simplices in the Calabi-Yau triangulation containing $\alpha_9$
(associated with $p_1$), say $\{\alpha_1,\ldots,
\alpha_7, \alpha_9,\ldots, \alpha_{12}\}$ (corresponding to the $x_8 \ne 0$
patch in the $\IP^8$) and $\{\alpha_1, \ldots, \alpha_6, \alpha_8,
\ldots, \alpha_{12}\}$ (corresponding to $x_7 \ne 0$), each of which has
volume 1. We now partially star unsubdivide on the $\alpha_9$ (i.e. remove
$\alpha_9$ from both simplices and form a resulting simplex as a union of
the remaining subsimplices), giving a nonzero expectation value to $p_1$
and obtaining a volume 2 resultant simplex formed by the blowing down of
a $\IP^1$ ($x_7 \ne 0$ or $x_8 \ne 0$). Thus we have collapsed 2 patches
of $\IP^8$ (whose union is the whole of $\IP^8$) into a singular point
over a $\IP^3$ patch ($p_1 \ne 0, p_2 \ldots p_4$ arbitrary). By doing
similar unsubdivisions for the other simplices we obtain a $\IC^8/\IZ_2$
fiber over a $\IP_3$ as expected.

An important observation, originally made in
 \ref\rBerglund{P. Berglund, P. Candelas, X. de la Ossa, A. Font, T.
Hubsch, D. Jan\v{c}i\'{c} and F.  Quevedo, {\it Periods for Calabi-Yau
and Landau-Ginzburg Vacua}, hep-th/9308005} is that the mirror to this
manifold, which has $h^{2,1} = 1$, admits a $\BZ_2$ symmetry on its
complex structure moduli space. Reinterpreted back on the original
manifold being studied here this identifies the two phases just discussed
\rAG. Thus, the ``small'' radius phase, involving a Landau-Ginzburg
fibration on a $\IP^3$, can be reinterpreted as a large radius Calabi-Yau
sigma model. This gives an unambiguous $R \rightarrow 1/R$ type symmetry
in the Calabi-Yau context.

\leftline{\tenbf Example 4}

	We now come to an example where all phase regions turn out to be
Calabi-Yau. Let's in particular consider the following example, which will
also contain a rather explicit symmetry relating the 4 phase regions.

\eqn\ecicyttt{\left(\matrix 2\cr2\cr2\cr \endmatrix\right|\left|
	\matrix 1&1&1\cr1&1&1\cr1&1&1\cr \endmatrix\right)}

	The point set here is
$${\offinterlineskip \lineskiplimit =1pt
	\eqalign{\alpha_1&=(1,0,0,0,0,0,1,0,0)\cr
	\alpha_2&=(0,1,0,0,0,0,0,1,0)\cr
	\alpha_3&=(-1,-1,0,0,0,0,0,0,1)\cr
	\alpha_4&=(0,0,1,0,0,0,1,0,0)\cr \alpha_5&=(0,0,0,1,0,0,0,1,0)\cr
	\alpha_6&=(0,0,-1,-1,0,0,0,0,1)\cr
	\alpha_7&=(0,0,0,0,1,0,1,0,0)\cr \alpha_8&=(0,0,0,0,0,1,0,1,0)\cr
	\alpha_9&=(0,0,0,0,-1,-1,0,0,1)\cr
	\alpha_{10}&=(0,0,0,0,0,0,1,0,0)\cr
	\alpha_{11}&=(0,0,0,0,0,0,0,1,0)\cr
	\alpha_{12}&=(0,0,0,0,0,0,0,0,1)}}$$

	There are only 4 triangulations of the above set of points, and
they are easily identified by the set of points that belong to every
simplex. The original Calabi-Yau has $\{\alpha_{10}, \alpha_{11},
\alpha_{12}\}$ in each of it's simplices. The other triangulations are
obtained from permuting this set. The superpotential here is nicely
symmetric in each set of variables. The D-term appears to treat the
$\phi$ and $p$-fields asymmetrically, however, the importance of this
asymmetry is only to determine which fields will span the direct sum of
line bundles over the product of $\IP^2$'s in which the Calabi-Yau is
embedded. The symmetry we are referring to  is linked  to the
choice of representation of the asymmetry, i.e. to the choice of fields
spanning the line bundles. To see this explicitly, we note that
minimization of the D-term requires $\sum_{i=1}^{3} |p_{i}|^2 =
\sum_{i=1}^{3} |\phi_{i}|^2 - r_1$. This allows us to substitute for the
sum of the $p_{i}'s$ in the other two parts of the D-term with $\phi_{1},
\phi_{2}, \phi_{3}$, leading to a D-term which has $\phi_1,\phi_2,
\phi_3$ as $p$-fields and 3 independent $r$-parameters $-r_1, r_2 - r_1,
r_3 - r_1$. Thus, all four phases are smooth CY and are actually
isomorphic.

\vglue0.6cm
\leftline{\tenbf 6. Conclusions}
\vglue0.4cm

The moduli space of N = 2 superconformal field theories contains within
it a surprising wealth of both physical and geometrical content. We have seen
that the phase diagram of those theories smoothly connected to one another
by truly marginal deformations  decomposes into numerous
phases. Within each phase there is a region for which perturbative methods
are reliable provided one expands around an appropriate description of
the physical model. The identity of such a model is thereby naturally
associated with each region.

Some time ago it was shown that there are examples which have regions
corresponding to Calabi-Yau sigma models on topologically distinct target
spaces.  Nonetheless, the physical model can smoothly interpolate from
one region to any other and hence the target space topology can change
without a physical discontinuity.  We described some further properties
of these phase diagrams in this talk and found some other interesting
features. We have seen that the sheer number of phase regions grows quite
quickly with the number of moduli parameters.  The``typical" such phase
diagram has regions identifiable as smooth Calabi-Yau sigma models,
orbifold Calabi-Yau sigma models, Landau-Ginzburg, and hybrid combinations
thereof. We have emphasized, though, that there are other examples in
which some of these typical types of models do not arise.  For instance,
we have seen examples where there is no variety of Landau-Ginzburg theory
-- all regions having a Calabi-Yau like interpretation. We have seen
other examples that give us the first unambiguous $R \rightarrow 1/R$
symmetry.  And finally, we have seen examples which have no variety of
Calabi-Yau sigma model interpretation. The latter have provided us with a
satisfying perspective on what it means to have mirror symmetry in the
context of rigid Calabi-Yau manifolds.

An interesting and important area to pursue is the application of all of this
formalism to the setting of $(0,2)$ models. Such work is in progress and
will be reported upon shortly.

\listrefs

\bye